\documentclass[fleqn,twoside]{article}
\usepackage{espcrc2}

\usepackage{graphicx}
\usepackage[figuresright]{rotating}

\newcommand{\AmS}{{\protect\the\textfont2
  A\kern-.1667em\lower.5ex\hbox{M}\kern-.125emS}}

\newcommand{\no}{\nonumber}

\title{Polarization effects in tau production by neutrino%
       \thanks{presented by K. Mawatari in oral session and by
       H. Yokoya in poster session
        at the 3rd International Workshop on Neutrino-Nucleus Interactions
        in the Few GeV Region (NuInt04), March 17-21, 2004, Gran Sasso,
        Italy.}}

\author{K. Hagiwara\address{Theory Group, KEK, Tsukuba 305-0801, Japan},
        K. Mawatari\address{Graduate School of Science and
                   Technology, Kobe University, Nada, Kobe 657-8501, Japan}%
           \thanks{email address: mawatari@radix.h.kobe-u.ac.jp} and
        H. Yokoya\address{Department of Physics, Hiroshima University,
                  Higashi-Hiroshima 739-8526, Japan}%
                 \address{Radiation Laboratory, RIKEN,
                  Wako 351-0198, Japan}%
           \thanks{email address: yokoya@theo.phys.sci.hiroshima-u.ac.jp} 
       }
       
\begin{document}

\begin{abstract}
We studied polarization effects in tau production by neutrino-nucleon
 scattering. 
Quasi-elastic scattering, $\Delta$ resonance production and deep
 inelastic scattering processes are taken into account for the
 CERN-to-Gran Sasso projects.
We show that the tau produced by neutrino has high degree of
 polarization, and its spin direction depends non-trivially on the
 energy and the scattering angle of tau in the laboratory frame.  
\vspace{1pc}
\end{abstract}
\maketitle

\setcounter{footnote}{0}
\section{INTRODUCTION}
In the forthcoming neutrino experiments, $\nu_\tau$
appearance which is the positive evidence of $\nu_\mu\to \nu_\tau$
oscillation is the one of main subjects, as well as precise measurements
of neutrino oscillation parameters and unknown $\theta_{13}$ or CP
phase measurement. 
In 2005, ICARUS \cite{icarus} and OPERA \cite{opera} experiments using 
the CNGS beam \cite{cngs} will start to detect $\nu_\tau$ appearance.

$\nu_{\tau}$ should be detected through 
the $\tau$ production by charged current reactions off a nucleon target.
As we pointed out in the recent paper \cite{taupol}, the information 
on the spin polarization of $\tau$ produced by neutrino is essential 
to determine the $\tau$ production signal since the decay particle 
distributions depend crucially on the $\tau$ polarization.

We present polarization effects in $\tau$ production by neutrino-nucleon
 scattering via charged current. 
We consider Quasi-elastic scattering (QE), $\Delta$ resonance production 
(RES) and deep inelastic scattering (DIS) processes, which give dominant
 contributions in the CNGS beam energy region, in the laboratory frame
\cite{kuzmin}.

\section{KINEMATICS AND FORMALISM}

We show the physical regions of kinematical variables and give the
relation between the $\tau$ spin polarization vector and the spin 
density matrix of the charged current $\tau$ production process. 
In this report, we take account of only $\tau^-$ production.
You can find more details in Ref. \cite{taupol}.

\subsection{Kinematical region for each process}

We consider $\tau^-$ production for each process by neutrino
off a nucleon target in the laboratory frame; 
\begin{equation}
\nu_{\tau}(k) + N(p) \to \tau^-(k') + 
 \left\{\begin{array}{rl}
          N'(p')      & ({\rm QE})  \\
        \Delta\, (p') & ({\rm RES}) \\
        X\,(p')       & ({\rm DIS}) 
        \end{array}
 \right.
\end{equation}
The four-momenta are
\begin{eqnarray}
k \!\!\!&=&\!\! (E_{\nu},0,0,E_{\nu}),\no \\
p \!\!\!&=&\!\!  (M,0,0,0),\\
k'\!\!\!&=&\!\!  (E_{\tau},p_{\tau}\sin\theta,0,p_{\tau}\cos\theta),\no
\end{eqnarray}
and Lorentz invariant variables are defined as
\begin{eqnarray}
&&Q^2 = -q^2,  \quad  q= k-k',\\
&&W^2 =(p+q)^2.
\end{eqnarray}
Each process is distinguished by the hadronic invariant mass $W$: 
$W=M$ for QE, and $M+m_\pi <W<W_{\rm cut}$ for RES.  
$W_{\rm cut}$ is an artificial boundary between the RES and 
DIS ($W>W_{\rm cut}$) processes%
\footnote{In Ref. \cite{graczyk}, the modifications to the PDF's
\cite{bodek} are considered.}, and we take $W_{\rm cut}=1.6$ GeV.

\begin{figure}[h]
\begin{center}
\includegraphics[width=6cm]{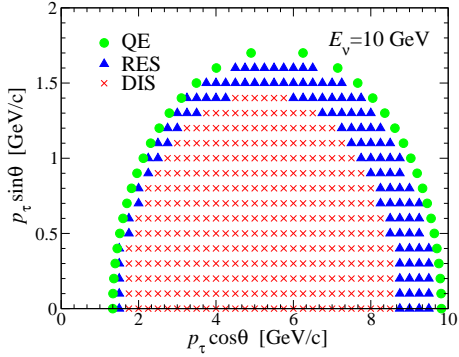}
\vspace*{-0.75cm}
\caption{Physical region at $E_\nu =10$ GeV}\label{region}
\end{center}
\end{figure}

Figure \ref{region} shows the kinematical region of 
each QE, RES and DIS process on the 
($p_\tau\cos\theta$, $p_\tau\sin\theta$) plane
at $E_\nu =10$ GeV.
\subsection{Spin polarization vector of $\tau$}

The spin polarization vector of $\tau$ is defined as 
\begin{equation}
\vec s = \frac{P}{2}(\sin\theta_P\cos\varphi_P,
             \sin\theta_P\sin\varphi_P,\cos\theta_P) \label{vector}
\end{equation}
in the $\tau$ rest frame in which the z-axis is taken along it's
momentum direction in the laboratory frame.
$P$ denotes the degree of polarization.
$P=1 $ gives the fully polarized $\tau$, and $P=0$ gives unpolarized $\tau$. 
The degree of polarization ($P$) and the spin directions 
($\theta_P, \varphi_P$) are functions of the tau energy $E_\tau$ 
and the scattering angle $\theta$.

This spin polarization vector is related to 
 the spin density matrix $R_{\lambda\lambda'}$:
\begin{equation}
\frac{dR_{\lambda\lambda'}}{dE_\tau\,d\cos\theta}
=(\frac{1}{2}+\vec s \cdot\vec\sigma)\,
\frac{d\sigma_{\rm sum}}{dE_\tau\,d\cos\theta}.  
\end{equation}
The spin density matrix is calculated as
$R_{\lambda\lambda'}\propto M_\lambda{M_{\lambda'}}^*$, 
where $M_\lambda$ is the helicity amplitude with the $\tau$ 
helicity $\lambda/2$ defined in the laboratory frame, and 
$d\sigma_{\rm sum}= dR_{++}+dR_{--}$ is the usual spin summed 
cross section.

The spin density matrix of $\tau$ production 
is obtained by using the leptonic and hadronic tensor as
\begin{equation}
\frac{dR_{\lambda\lambda'}}{dE_\tau\, d\cos\theta}
= \frac{G_F^2\kappa^2}{4\pi} \frac{p_\tau}{ME_\nu}\,
  L^{\mu\nu}_{\lambda\lambda'}W_{\mu\nu},
\end{equation}
where $G_F$ is Fermi constant and $\kappa=M_W^2/(Q^2+M_W^2)$. 
The leptonic tensor is expressed as 
\begin{equation}
L_{\lambda\lambda'}^{\mu\nu}=j_\lambda^\mu\,{j^\nu_{\lambda'}}^*,
\end{equation}
where the leptonic weak current $j_\lambda^\mu$ is 
\begin{equation}
j_\lambda^\mu= \bar u_\tau(k',\lambda)\,\gamma^\mu 
               \frac{1-\gamma_5}{2}\, u_\nu (k). 
\end{equation}

\section{HADRONIC TENSORS}

We show the hadronic tensors which we used in our calculation 
for each process, i.e., QE, RES and DIS. 
Several parameters and form factors were updated to the recent our paper
\cite{taupol}.

\subsection{Quasi-elastic scattering (QE)}

The hadronic tensor for the QE scattering process;
\begin{equation}
\nu_{\tau} + n \to \tau^- +p
\end{equation}
is written by using the hadronic weak transition current
$J_\mu$ as follows \cite{llewellyn}:
\begin{equation}
W^{\rm QE}_{\mu\nu}=\frac{\cos^2\theta_c}{4}\sum_{\rm spins}
J_\mu{J_\nu}^*\,\delta(W^2-M^2),
\end{equation}
where $\theta_c$ is the Cabibbo angle. $J_\mu$ are defined as
\begin{equation}
J_\mu=\bar u_p(p')\,\Gamma_\mu\,u_n(p),
\end{equation}
where $\Gamma_\mu$ is written in terms of the six weak form factors 
of the nucleon, $F^V_{1,2,3}$, $F_A$, $F_3^A$ and $F_p$, as
\begin{eqnarray}
\Gamma_\mu \!\!\!\!&=&\!\!\!\! 
  \gamma_\mu F^V_1 
 +\frac{i\sigma_{\mu\alpha}q^\alpha\xi}{2M}F^V_2
 +\frac{q_\mu}{M}F^V_3 \no\\
\!\!\!&&+\! 
  \left[ \gamma_\mu F_A
 +\frac{(p+p')_\mu}{M}F^A_3
 +\frac{q_\mu}{M}F_p \right]\gamma_5.
\end{eqnarray}
We can drop $F^V_3$ and $F^A_3$ because of time reversal invariance 
and isospin symmetry. Moreover, the vector form factors
$F_1^V$ and $F_2^V$ are related to the electromagnetic form factors 
of nucleons under the conserved vector current hypothesis:
\begin{eqnarray}
&&F^V_1(q^2)= 
 {G^V_E-\frac{q^2}{4M^2}G^V_M \over 1-\frac{q^2}{4M^2}}, \no \\
&&\xi F^V_2(q^2)=\frac{ G^V_M-G^V_E}{ 1-\frac{q^2}{4M^2}},
\end{eqnarray}
where 
\begin{eqnarray}
G^V_E=\frac{1}{(1-q^2/M_V^2)^2},\ 
G^V_M=\frac{1+\xi}{(1-q^2/M_V^2)^2},
\end{eqnarray}
with a vector mass $M_V=0.84$ GeV and $\xi=3.706$. 
For the axial form factor $F_A$ and $F_p$, we use:
\begin{eqnarray}
&&F_A(q^2)=\frac{F_A(0)}{(1-q^2/M_A^2)^2}, \\
&&F_p(q^2)=\frac{2M^2}{m_\pi^2-q^2}F_A(q^2),
\end{eqnarray}
with $F_A(0)=-1.27$ \cite{PDG} and an axial vector mass $M_A=1.026$ 
GeV \cite{bernard}.
Notice that the pseudoscalar form factor $F_p$ plays an important 
role for the polarization of $\tau$ produced by neutrino
because its contribution is proportional to the lepton mass and 
it has the spin-flip nature, although it is not known well.
In Ref. \cite{psff}, we discussed it in detail.

\subsection{Resonance production (RES)}
The hadronic tensor for the $\Delta$ resonance production (RES) process;
\begin{equation}
\nu_{\tau}+n\ (p) \to \tau^-+\Delta^+\ (\Delta^{++})
\end{equation}
is calculated in terms of the nucleon-$\Delta$ weak transition current 
$J_{\mu}$ as follows \cite{llewellyn,schreiner,singh}:
\begin{eqnarray}
W^{\rm RES}_{\mu\nu}=
\frac{\cos^2\theta_c}{4}\sum_{\rm spins}J_\mu {J_\nu}^*\, |f(W)|^2,
\end{eqnarray}
where $f(W)$ is the Breit-Wigner factor
\begin{equation}
 f(W)={\sqrt{W\Gamma (W)/\pi} \over W^2-M_{\Delta}^2+iW \Gamma (W)}
\end{equation}
with its running width
\begin{equation}
\Gamma(W) = 
 \Gamma(M_{\Delta})\,\frac{M_{\Delta}}{W}\,\frac{\lambda^{\frac{1}{2}}
(W^2,M^2,m_{\pi}^2)}{\lambda^{\frac{1}{2}}(M_{\Delta}^2,M^2,m_{\pi}^2)}. 
\end{equation}
$\Gamma(M_{\Delta})=0.12$ GeV and $\lambda(a,b,c)=a^2+b^2+c^2-2(ab+bc+ca)$.
The current $J_{\mu}$ for the process 
$\nu_{\tau}+n\to\tau^{-}+\Delta^{+}$ is defined by 
\begin{equation}
J_{\mu}=\langle\Delta^{+}(p')|\hat{J}_{\mu}|n(p) \rangle 
=\bar{\psi}^{\alpha}(p')\,\Gamma_{\mu\alpha}\,u_{n}(p),
\end{equation}
where $\psi^{\alpha}$ is the spin-3/2 particle wave function and 
the vertex $\Gamma_{\mu\alpha}$ is expressed in terms of 
the eight weak form factors 
$C^{V,A}_{i=3,4,5,6}$ as 
\begin{eqnarray}
\Gamma_{\mu\alpha}\!\!\!\!&=&\!\!\!\!\left[
\frac{g_{\mu\alpha}\!\not\!q-\gamma_{\mu}q_{\alpha}}{M} C^V_3
+\frac{g_{\mu\alpha}\,p'\cdot q-p'_{\mu}q_{\alpha}}{M^2} C^V_4 
\right. \no \\ 
\!\!\!\!&+&\!\!\!\! \left. 
\frac{g_{\mu\alpha}\, p\cdot q - p_{\mu}q_{\alpha}}{M^2} C^V_5
+\frac{q_{\mu}q_{\alpha}}{M^2} C^V_6 \right]\gamma_5 \no\\
\!\!\!\!&+&\!\!\!\! 
\frac{g_{\mu\alpha}\!\not\!q -\gamma_{\mu}q_{\alpha}}{M} C^A_3
+\frac{g_{\mu\alpha}\,p'\cdot q -p'_{\mu}q_{\alpha}}{M^2} C^A_4 
\no \\
\!\!\!\!&+&\!\!\!\! g_{\mu\alpha} C^A_5
+\frac{q_{\mu}q_{\alpha}}{M^2}  C^A_6.
\end{eqnarray}
By using the isospin invariance and the Wigner-Eckart theorem, 
we obtain another nucleon-$\Delta$ weak transition current as 
\begin{equation}
\langle\Delta^{++}|\hat{J}_{\mu}|p \rangle = 
\sqrt{3}\langle\Delta^{+}|\hat{J}_{\mu}|n \rangle .
\end{equation}
From the CVC hypothesis, $C^{V}_{6}=0$ and the other vector 
form factors $C^{V}_{i=3,4,5}$ are related to 
the electromagnetic form factors. We adopt the modified dipole  
parameterizations \cite{pys,olsson}:
\begin{eqnarray}
&& C^V_3(q^2)=\frac{C_3^V(0)} {(1-{q^2\over M_V^2})^2}
{ 1\over 1- { q^2\over 4M_V^2}}, \no \\
&& C^V_4(q^2)=-\frac{M}{M_{\Delta}}C^V_3(q^2),\quad
 C^V_5(q^2)=0,
\end{eqnarray}
with $C_3^V(0)=2.05$ and $M_V=0.735$ GeV. 
For axial form factors, we use \cite{pys}
\begin{eqnarray}
&&C^A_5(q^2) = {C_5^A(0) \over (1- {q^2\over M_A^2})^2} 
    {1 \over 1-{q^2 \over 3M_A^2}}, \\
&&C^A_6(q^2)=\frac{M^2}{m_{\pi}^2-q^2} C_5^A(q^2),
\end{eqnarray}
with $C_3^V(0)=2.05$ and $M_A=1.0$ GeV.
 For $C_3^A$ and $C_4^A$, $C_3^A=0$ and $C_4^A=-{1\over 4}C_5^A$ give 
good agreements with the data \cite{schreiner}.
As in the case of the $F_{p}(q^{2})$ of the QE process, 
$C_{6}^{A}(q^{2})$ has significant effects on the $\tau$ production 
cross section and the $\tau$ polarization \cite{psff}.
\begin{figure*}[ht]
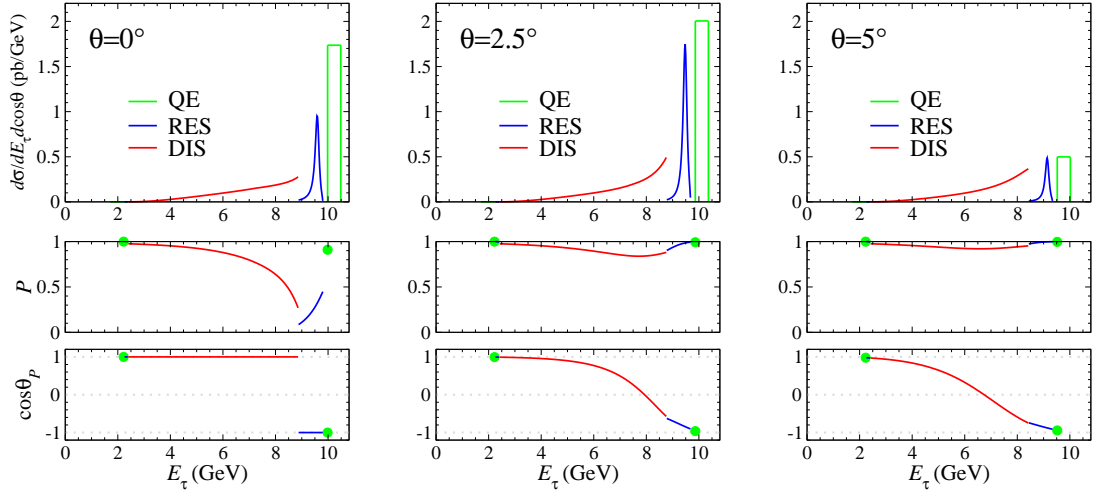

\begin{center}
\includegraphics[height=6.5cm]{0d.eps}\hspace*{0.62cm} 
\includegraphics[height=6.5cm]{2d.eps}\hspace*{0.62cm}
\includegraphics[height=6.5cm]{5d.eps}
\end{center}
\vspace*{-0.8cm}
\caption{
Production cross section and the $\tau$ polarization of the process 
${\nu}_{\tau}N \to \tau^{-}X$ at $E_{\nu}=10$ GeV. 
$E_{\tau}$ dependence 
of the differential cross section (top), the degree of 
polarization $P$ (middle) and the polar component of the normalized 
polarization vector $\cos\theta_P$ (bottom) are shown along the laboratory 
frame scattering angle $\theta=0^{\circ}$ (left), $2.5^{\circ}$ (center) and
$5^{\circ}$ (right), respectively. 
}\label{taupol}
\end{figure*}

\subsection{Deep inelastic scattering (DIS)}

In the DIS region, the hadronic tensor is estimated by using the
quark-parton model;
\begin{equation}
W^{\rm DIS}_{\mu\nu}(p,q)=\!\sum_{q,\bar q}\!\!\int\!\!\frac{d\xi}{\xi}
f_{q,\bar q}(\xi,Q^2) K^{(q,\bar q)}_{\mu\nu}(p_q,q),
\end{equation}
where $p_q=\xi p$ is the four-momentum of the scattering quark, 
$\xi$ is its momentum fraction, and  
$f_{q,\bar q}$ are the parton distribution function(PDF)'s 
inside a nucleon. The quark tensor $K^{(q,\bar{q})}_{\mu\nu}$ is 
\begin{eqnarray}
K^{(q,\bar q)}_{\mu\nu} \!\!\!\!\!\!\!&(&\!\!\!\!\!\!\!p_q,q)=
\delta(2\,p_q\cdot q-Q^{2}-m_{q'}^2)\no \\
&\times&\!\!\!\!\!\! 2[-g_{\mu\nu}(p_q\cdot q)+2p_{q\mu}p_{q\nu} \no \\
&& + p_{q\mu}q_\nu +q_\mu p_{q\nu}
\mp i\epsilon_{\mu\nu\alpha\beta}p_{q}^\alpha q^\beta].
\end{eqnarray}
The upper sign should be taken for quarks and the lower for antiquarks. 
We retain the final quark mass $m_{q'}$ for the charm quark as  
$m_{c}=1.25$ GeV, but otherwise we set $m_{q'}=0$. 
In the calculation, we used the MRST2002 \cite{mrst} as the PDF's%
\footnote{We use naive extrapolation of the parton model calculation, even when 
$Q^2<Q^2_0$ (=1.25 GeV$^2$ in MRST2002) }.

 \section{POLARIZATION OF PRODUCED $\tau^-$}

We show the spin polarization vector 
of $\tau$ lepton produced by neutrino off isoscalar targets as a 
function of its 
energy $E_{\tau}$ and the scattering angle $\theta$ in the laboratory frame. 

Figure \ref{taupol} summarizes our results%
\footnote{The Fortran code for our calculation of tau polarization 
is available on the web \cite{hiroshi}.} for the 
${\nu}_{\tau}N \to \tau^{-}X$ process at $E_{\nu}=10$ GeV.

The top three figures show the double differential cross 
sections as a function of $E_{\tau}$  at the scattering angle  
$\theta=0^{\circ}$ (left figures), $5^{\circ}$ (center figures) and
$10^{\circ}$ (right figures).
The area of the histogram for the QE process is normalized to the cross 
section. 
The QE and RES cross sections are large at 
forward scattering angles, and the DIS contribution become more significant 
at large scattering angles, though the cross section gets smaller.

A set of three middle figures give the degree of polarization $P$, and 
in the bottom three figures we show the polarization direction
$\cos\theta_P$ in Eq. (\ref{vector}). 
The produced $\tau^-$ is almost fully polarized except at 
 the very small scattering angle. 
As for the angle of the polarization vector, 
the high energy $\tau^-$ is almost left-handed ($\cos\theta_P=-1$). 
On the other hand, the spin of low energy $\tau^{-}$ turns around. 
The azimuthal angle $\varphi_{P}$ takes $\pi$ at all energies, 
which means that the spin vector points to the direction of 
the initial neutrino momentum axis.

In order to understand the above features, it is useful to consider the
polarization of $\tau^-$ in the center of mass (CM) frame of the 
scattering particles. Let us consider the DIS process in the $\nu q$ 
CM frame, since the $\nu q$ scattering is dominant in the 
$\nu_{\tau}N\to\tau^{-}X$ process. In this frame, 
produced $\tau^{-}$ is fully left-handed polarized at all scattering angles. 
This is because the initial $\nu_{\tau}$ and $q$ ($d$ or $s$ quarks) are 
both left-handed and hence angular momentum along the initial momentum 
direction is zero, while in the final state the produced $u$ quark is 
left-handed and hence only the left-handed $\tau^-$ is allowed by the angular 
momentum conservation. This selection rule is violated slightly when a 
charm quark is produced in the final state and because of gluon radiation 
at higher orders of QCD perturbation theory. The $\tau^-$ polarization in the 
laboratory frame is then obtained by the Lorentz boost.
In the QE and RES processes, situations are
almost the same as in the DIS process. In the CM frame of $\nu N$ collisions, 
the $\tau^-$ lepton produced by the QE or RES process is almost 
left-handed at all angles, for the CM energy of 
$\sqrt{2ME_{\nu}+M^2}\approx 4.4$ GeV for $E_{\nu}=10$ GeV, for our 
parametrizations of the transition form factors.
High energy $\tau^-$'s in the laboratory frame have 
left-handed polarization because those $\tau^-$'s have forward scattering 
angles also in the CM frame.
However, lower energy $\tau^-$'s in the laboratory frame tends to have 
right-handed polarization because they are produced at 
backward angles in the CM frame. 
At the zero scattering angle $\theta=0^{\circ}$ of the laboratory frame, 
the change in the $\tau^-$ momentum direction occurs suddenly, 
and hence the transition from the left-handed $\tau^-$ at high energies 
to the right-handed $\tau^-$ at low energies is discontinuous.

\section{CONCLUSION}

The information on the polarization of $\tau$ produced 
through the neutrino-nucleon scattering is 
essential to identify 
the $\tau$ production signal since the decay particle distributions 
depend crucially on the $\tau$ spin polarization. 
It is needed in long baseline neutrino oscillation experiments
which should verify the large $\nu_{\mu}\to \nu_{\tau}$ 
oscillation, and is also needed for the background estimation of 
$\nu_{\mu}\to\nu_{e}$ appearance experiments which should measure 
the small mixing angle of $\nu_{e}$-$\nu_{\mu}$ oscillation. 

In this report we presented the spin polarization of 
$\tau^-$ produced in tau-neutrino nucleon scattering via charged currents. 
Quasi-elastic scattering (QE), $\Delta$ resonance production (RES) and
deep inelastic scattering (DIS) processes have been studied.
The three subprocesses are distinguished by the hadronic invariant mass $W$. 
$W=M(=m_N)$ gives QE, $M+m_{\pi}<W<W_{\rm cut}$ gives RES, and 
$W>W_{\rm cut}$ gives DIS. 
Here we set the kinematical boundary of RES and DIS process
at $W_{\rm cut}=1.6$ GeV. 

The spin density matrix of $\tau$ production 
has been defined and the $\tau$ spin polarization vector has been
defined and parametrized in the $\tau$ rest frame whose polar-axis 
is taken along the momentum direction of $\tau$ in the laboratory frame. 
The spin density matrix has been calculated for each subprocess by using the 
form factors for the QE and RES processes, and by using the parton 
distribution functions for the DIS process.
We have shown the spin polarizations of $\tau$ as function of the 
$\tau$ energy and the scattering angle in the laboratory frame for 
$\nu_{\tau}N \rightarrow \tau^-X$ process at $E_{\nu}=10$ GeV.
We found that the produced $\tau^-$ have 
high degree of polarization, but their spin directions deviate significantly 
from the 
massless limit predictions at low and moderate $\tau$ energies.
Qualitative feature of the predictions have been 
understood by considering the helicity amplitudes 
in the CM frame of the scattering particles and the effects of Lorentz 
boost from the CM frame to the laboratory frame. 

\begin{figure}[h]
\vspace*{-0.2cm}
\begin{center}
\includegraphics[width=6.5cm]{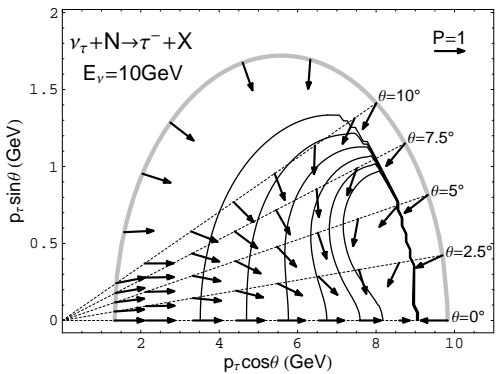}
\end{center}
\vspace*{-1cm}
\caption{The contour map of the DIS cross section in the 
($p_{\tau}\cos\theta$, $p_{\tau}\sin\theta$) plane for the  
$\nu_{\tau}N\to \tau^{-}X$ process at $E_{\nu}=10$ GeV in the laboratory frame.
The kinematical boundary is shown by the thick gray curve. The QE process 
contributes along the boundary, and the RES process contributes just inside 
of the boundary. The $\tau^-$ polarization are shown by the arrows. 
The length of the arrows give the degree of
 polarization, and the direction of arrows give that of the $\tau^{-}$
spin in the $\tau^{-}$ rest frame. The size of the 100\% polarization ($P=1$) 
arrow is shown as a reference. The arrows are shown along the laboratory 
scattering angles, $\theta=0^{\circ}$, $2.5^{\circ}$, $5^{\circ}$, 
$7.5^{\circ}$, and $10^{\circ}$, as well as along the kinematical boundary.  
}\label{convec}
\end{figure}

Finally, we summarize our findings in Figure \ref{convec}. 
In Figure \ref{convec}, we show the polarization vector $\vec{s}$ of $\tau^{-}$
for the $\nu_{\tau}N \rightarrow \tau^-X$ process at $E_{\nu}=10$ GeV 
on the ($p_{\tau}\cos\theta$, $p_{\tau}\sin\theta$) plane, 
where $p_{\tau}$ and $\theta$ are the produced $\tau$ momentum 
and the scattering angle in the laboratory frame. 
The length of each arrow gives the degree of 
polarization ($0\leq P \leq 1$) at each phase-space point and its 
orientation gives the spin direction in the $\tau^-$ rest frame. 
The differential cross section is described as a contour map, 
where only the DIS cross section is plotted to avoid too much complexity.
The outer line gives the kinematical boundary, along 
which the QE process occurs.
Figure \ref{convec} is a more visual version of the information given in 
Figure \ref{taupol}.
\section*{Acknowledgements}

K.M. and H.Y. are grateful to the organizers for the interesting and
pleasant workshop and also for giving us the opportunity to present
this work. 
K.M. would like to thank T. Morii and M. Sakuda for encouragements.
H.Y. would like to thank RIKEN BNL Research Center for the support of
travel expense for this workshop.

\end{document}